\documentclass[twocolumn,prl,amsmath,floatfix]{revtex4}
\usepackage{graphicx,color}

\newcommand{\figone}{\begin{figure}
{\includegraphics[width=3.25in,clip]{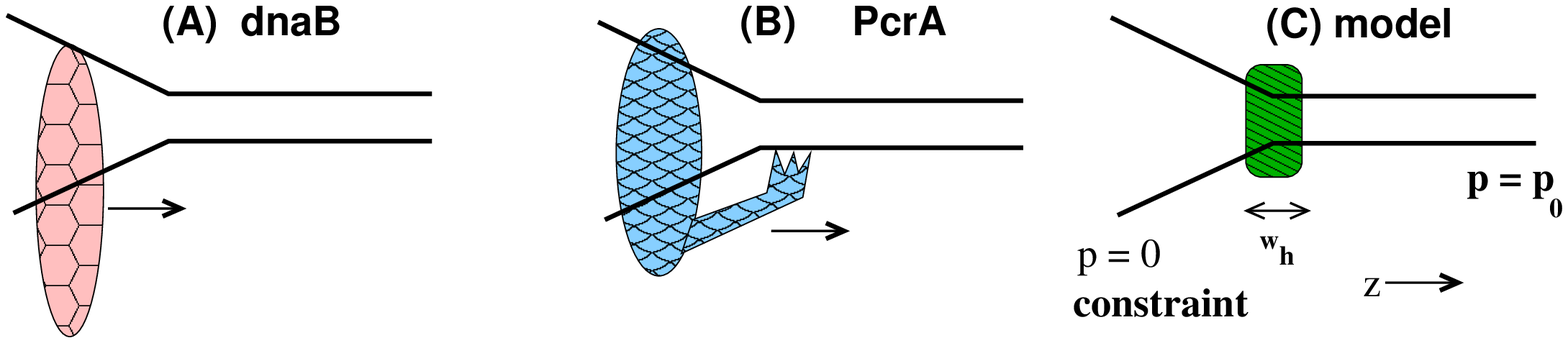}}
\caption{ Schematic representation of the advocated mechanisms
  for (A) dnaB and (B) PcrA helicases (see text) whose motions are
  indicated by the arrows.  (C) Proposed model.  Constraints are that
  the DNA is open on one side but zipped on the other, both phases
  coexisting.  The unzipped phase is preferred within the shaded box
  of width $w_h$ encompassing the Y-fork or interface or domain-wall.
}
\label{fig:hel}
\end{figure}
}

\newcommand{\figtwo}{
\begin{figure}
{\includegraphics[width=3.2in,clip]{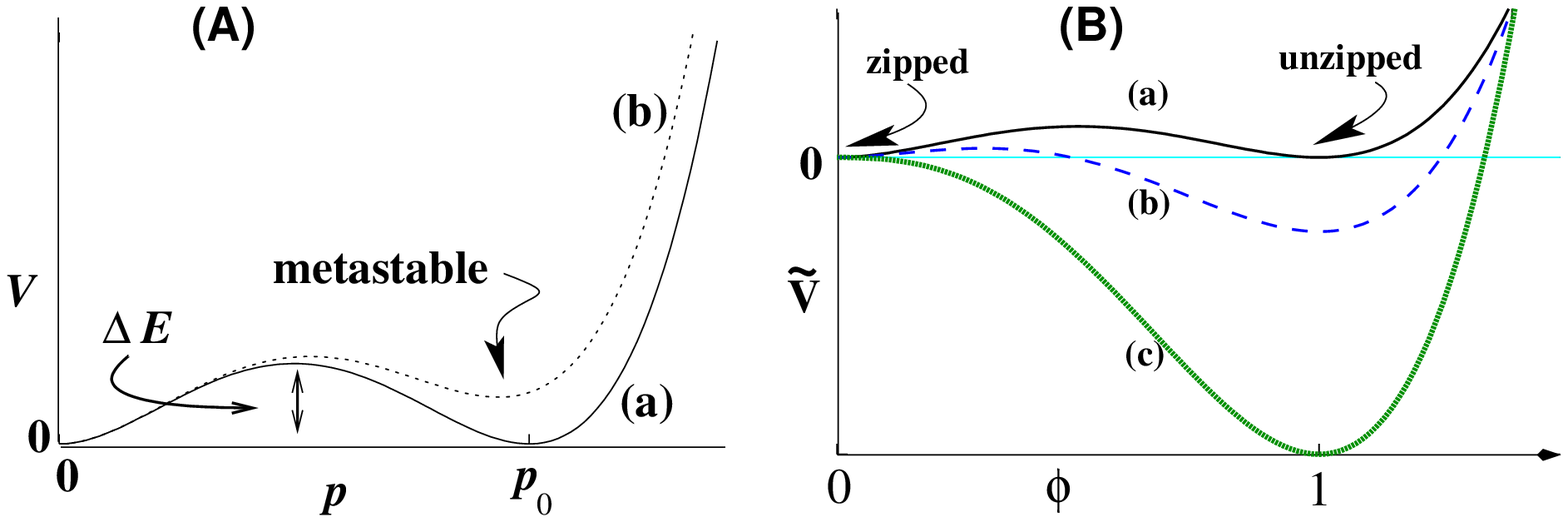}}
\caption{ Plots of  (A) $V(p)$  and (B)
  $\tilde{V}(\phi)$.  Curves (a) 
  show the co-existence of the zipped ($p= p_0$ or $\phi=0$) and the
  unzipped ($p=0$ or $\phi=1$) phases.  Curves (b) and (c) show the
  change near the interface or domain wall due to the drive ($h$ in
  Eq.  (\ref{eq:18})) that makes the zipped phase locally metastable
  or unstable.  }
\label{fig:res}
\end{figure}
}

\newcommand{\figthree}{
\begin{figure}
{\includegraphics[width=3.2in,clip]{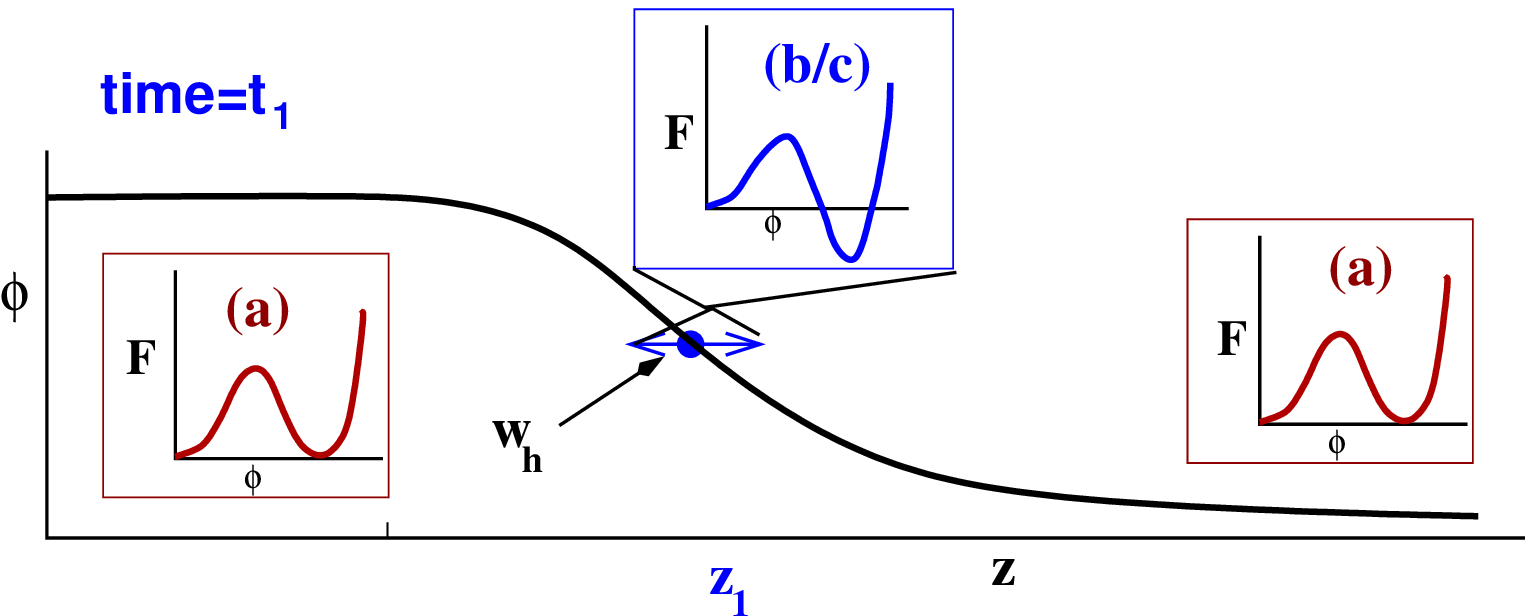}}
\caption{ Schematic representation of the pulse:  
  The front is at location $z_1$ (filled circle at $\phi=0.5$) at an
  arbitrary time $t_1$.  The effective free energy is of form (b) or
  (c) of Fig. \ref{fig:res} (metastable or unstable) only within
  $z_1\pm w_h$.  The drive or pulse moves with the front.  The
  relaxation due to the elastic term leads to a traveling wave
  solution. }
\label{fig:pul}
\end{figure}
}

\newcommand{\figfour}{\begin{figure}
{\includegraphics[width=2.5in,clip]{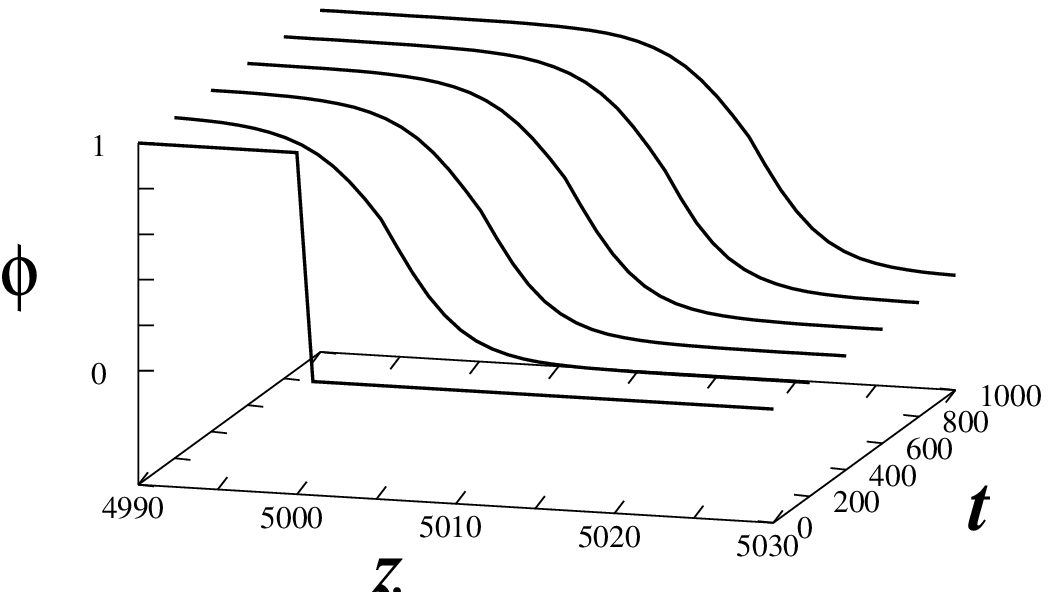}}
\caption{ Positions of the front or domain-wall
  for $\Delta u=0.2$ (see below Eq. \ref{eq:4}) at bulk coexistence
  are shown by plotting 
  $\phi(z)$ vs $z$.  The curves are at times $0, 200,400,600,800,
  1000$ in arbitrary time units and are obtained from a discretized
  version of Eq. (\ref{eq:18}).}
\label{fig:prob}
\end{figure}
}

\newcommand{\figfive}{\begin{figure}
{\includegraphics[width=2.5in,clip]{vel_wid_xmgr.eps}}
\caption{ Plot of $c(\Delta u,w_h)$ as a function of the width $w_h$ of
  Eq. (\ref{eq:4}) for different values of $\Delta u$ as noted
  against the curves.  Solid lines are fits to $c(\Delta u,w)= c_0[1 -a
  \exp(-w/w_0)]$.  }
\label{fig:vel}

\end{figure}
}
\begin{document}
 
\title{Helicase activity on DNA as a  propagating front}
\author{ Somendra M. Bhattacharjee}
\email{email: somen@iopb.res.in }
\affiliation{Institute of Physics, Bhubaneswar 751 005, India}

\begin{abstract}
  We develop a propagating front analysis, in terms of a local
  probability of zipping, for the helicase activity of opening up a
  double stranded DNA (dsDNA).  In a fixed-distance ensemble
  (conjugate to the fixed-force ensemble) the front separates the
  zipped and unzipped phases of a dsDNA and a drive acts locally
  around the front.  Bounds from variational analysis and numerical
  estimates for the speed of a helicase are obtained.  Different types
  of helicase behaviours can be distinguished by the nature of the
  drive.
\end{abstract}
\date{\today}   
\maketitle

A helicase moves along a DNA and unwinds it.  Whenever a double
stranded DNA in a cell needs to be opened up, a helicase is summoned,
be it during the semi-conservative replication, repair mechanism of a
stalled process or other DNA related activities\cite{biobook}.  
A large number of helicases including rna-helicases  have so far been
identified from different pro- and eucaryotes. A well-studied bacterium
like E. Coli contains at least 17 different helicases, though the need
and the function of each of these are not yet clear.

The helicase activity involves a motor action fed by NTP's (nucleotide
triphosphate) and eventual opening of DNA by dissociating successive
base pairs along the chain\cite{helicaserev}.  Quantitative estimates
of rates of such activities ($\sim 400$ base pair per second or less)
for almost all helicases are known from {\it in-vitro} studies in
solutions and more recently from single molecular experiments.
Attempts to categorize these varieties of helicases as per their
common features have led to various classification schemes.  These
are: {\it (i)} active vs passive depending on the direct requirement
of NTP for the opening; {\it (ii)} families and superfamilies (SF)
based on the conserved motifs of the primary sequence; {\it (iii)}
monomeric, dimeric, hexameric, oligomeric  depending on the number of
units required for activity; and {\it (iv)} mode of translocation:
whether it translocates on the single stands of DNA or on dsDNA.  For
example, dnaB, the main helicase involved in the replication of DNA in
E. Coli, is a hexameric, passive helicase belonging to the dnaB-like
family translocating on a single strand DNA\cite{dnaB}.  PcrA is an
active, SF1, monomeric helicase translocating on ssDNA\cite{pcra}
while recG is an active, SF2, monomeric, ds-DNA translocating
helicase\cite{recG}.  Apart from these gross classifications, very
little information is available on the detailed mechanism of the
helicase activity.

\figone

Crystallographic data  available for a few helicases have been
used to model mechanisms for specific helicases.  Though crystal
structures cannot give a dynamic view, such proposals, attractive no
doubt, are the only ones available so far.  According to these
proposals, hexameric helicase like dnaB, opens up dsDNA like a wedge
by virtue of its motor action on ssDNA\cite{dnaB}. A rolling mechanism
has been proposed for dimeric helicases\cite{dimerrolling}.  In case
of PcrA, the helicase activity and the motor action can be decoupled.
Crystal structure, supplemented by biochemical evidences on mutants, 
favours a mechanism where the helicase moves forward on ss-DNA and
during its sojourn a different domain of the helicase pulls a few
bases of a strand of the ds-DNA beyond the Y-fork, the junction
between ss and ds-DNA\cite{pcra}.  See Fig. \ref{fig:hel} for a
schematic diagram. RecG is more complicated because it moves in
opposite direction from zipped to unzipped, a fork reversal process
forming a Holliday junction of 4 dna strands\cite{recG}.

Our aim is to develop a generic physical picture that could be
applicable to all the different types of helicases.  Recently, a phase
coexistence based mechanism for helicase activity has been proposed by
Bhattacharjee and Seno\cite{hel2k2}.  A kinetic model has also been
proposed recently\cite{julicher} while a random walk model was used in
an earlier study\cite{chen} to analyze the movement on DNA.  The
phase-coexistence mechanism is based on the unzipping phase transition
of a ds-DNA by a force, which was first shown in a continuum model in
Ref. \cite{unzip}.  This transition has since then been established by
exact calculations for lattice models in all
dimensions\cite{unzip2,unzip3}, in studies of
dynamics\cite{sebastian,unzip2}, by scaling theories\cite{kafri}, in 
quenched averaged DNA \cite{nelson}, and others.  The key points of
the mechanism\cite{hel2k2} are the following. {\it (a)} A helicase
keeps two single strands of a dsDNA at a separation much bigger than
the base-pair separation of a ds-DNA.  {\it (b)} The resulting
fixed-distance ensemble for a dsDNA breaks it up into a zipped phase
and an unzipped phase separated by a domain wall.  {\it (c)} The
domain wall is identified as the Y-fork junction.  All helicases act
at or near this junction of the ds-ss DNA . {\it (d)} the motor action
of the helicase leads to a shift in the position of the fixed-distance
constraint thereby shifting the domain wall towards the zipped phase.
Additional features are needed for efficiency, job-requirement and
processivity (the distribution of the length unzipped before a
helicase drops off).  Our purpose in this paper is to use this
coexistence hypothesis to develop a simple coarse grained model of the
propagation of the Y-fork.

\figtwo

We use the zipping probability $p(z)$, the probability that the base
pair at index or contour length $z$ (measured along the backbone) is
zipped as the basic variable.  The unzipping transition by a force has
hitherto been studied by using polymer models.  However, in the case
of the conjugate fixed distance ensemble, the probability of zipping,
$p(z)$, under the imposed constraints, has been shown to be a useful
description\cite{mut2k2,hel2k2}.  In the polymer approach, one
introduces a local variable $n(z)=1$ or $0$ depending on whether the
base pairs at $z$ are bound or not to write the pairing energy,
characteristic of the DNA problem, as $\int dz \epsilon(z) n(z)$ where
$\epsilon(z)$ is the base pairing energy at $z$.  The average value
$\langle n(z)\rangle$ gives the fraction $p_0$ of bound basepairs, and
it is the parameter monitored in melting and unzipping
transitions.  This suggests the use of $p(z)$ as a coarse-grained
parameter to use for the inhomogeneous case, especially for the
description of the interface.  Since the unzipping transition is of
first order nature with a co-existence on the phase boundary, and our
interest is in the region much below the thermal melting point of DNA,
a phenomenological Landau-type hamiltonian or free energy can be used
to describe the state of the DNA.  In addition, the presence of the
helicase can be represented by the boundary conditions on two sides of
the DNA (zipped on one side and unzipped on the other).  The effect of
the motor action that pushes (or pulls) one phase into the other and
any other destabilizing effects of the helicase are taken into account
by additional terms involving the zipping probability $p(z)$.  We show
the existence of a traveling wave solution (i.e. a propagating front
of Y-fork) and then get the speed of the front in terms of the
parameters of the hamiltonian.  A variational principle is used to
derive bounds on the selected velocity.

The unzipping  transition can  be described by a
Landau-type hamiltonian or free energy
\begin{equation} 
 H_0 =
\int dz \left[ \frac{K}{2}\ \dot{p}(z)^2
+ V(p(z))\right ], 
\label{eq:1b}
\end{equation}
{\rm with} \ 
\[V(p) =
\frac{1}{2}\ r p^2 + \frac{1}{3}\ w
  p^3 + \frac{1}{4}\ u p^4, \]  
where $K$ is the appropriate rigidity modulus, $\dot{p}(z)=\partial
p(z)/\partial z$, and the cubic term
ensures a first order transition for $w<0, u>0$.  See Fig. \ref{fig:res}.
The coefficients $r, w$ and $u$ can in principle be determined from the
knowledge of the unzipping phase boundary and $p_0$.  It is often
useful (and used here also) to
reparametrize $V(p)$ in terms of $p_0$, the width of the well at $p_0$
and the barrier height $\Delta E$. 
Treating Eq. (\ref{eq:1b}) as the meanfield
free energy\cite{ebert}, the dynamics of unzipping is given by the
overdamped equation of motion
\begin{equation}
  \label{eq:5}
  \Gamma^{-1}\frac{\partial p(z)}{\partial t} = 
    -\frac{\delta H_0\ \ }{\delta p(z)} 
\end{equation}
with appropriate transport coefficient $\Gamma$.
Since we are away from critical points and interested in nonequilibrium 
propagation problem, we may ignore noise terms  in  the equation of
motion. Stochastic terms would also be required to describe
processivity, which we do not consider in this paper.

\figthree

For a long chain, with the boundary conditions $p(z) \rightarrow p_0$
(probability in the zipped phase) as $z\rightarrow +\infty$ and $p(z)
\rightarrow 0$ as $z\rightarrow -\infty$ at phase coexistence, the 
variation of the equilibrium probability of zipping is described by the 
equation $K \ddot{p}(z) = {V}^{\prime}(p)$, with prime (dot) denoting
derivative with respect to  the argument ($z$).
For the assigned boundary condition, there is a domain wall solution
located at an arbitrarily chosen $z=0$ with a profile $ \int_0^p
dp/\sqrt{2V(p)/K} = z,$ and $p(z)$ approaches the bulk value
exponentially in the zipped phase as $p_0-p(z) \sim
\exp(-V^{\prime\prime}(p_0)\mid z\mid/\sqrt{K})$.  The energy of a
wall of width $W$, in terms of the barrier height $\Delta E$
(Fig. \ref{fig:res}), is $E= K\frac{p_0^2}{W} + \Delta E \ W,$  which on
minimization gives $W = (K p_0^2/\Delta E)^{1/2}$.  Measurements of
$W$, and of $\Delta E$ (from the decay rate of bubbles in the bound
state) would give an estimate of $K$.  This equilibrium situation can
be obtained by keeping the helicase static on dsDNA, e.g. by denying
ATP in {\it in vitro} experiments.  Since such configurations can now
be prepared\cite{bianco}, detailed characterization of the wall can be
done experimentally.

In order to incorporate the effect of the motion of the helicase, we
introduce a moving perturbation that tends to destabilize the domain
wall at its current location.  A time dependent perturbation is
introduced in the equation of motion, Eq. (\ref{eq:5}) or,
equivalently, in the Hamiltonian of Eq. (\ref{eq:1b}), that favours
the unzipped state over a region of width $w_h$ around the domain
wall, maintaining co-existence elsewhere. See Figs. \ref{fig:res} and
 \ref{fig:pul}.  The crystal structure of
PcrA\cite{pcra} suggests $w_h \sim 20$ bases. 
The equation of motion, augmented by a ``drive'' term,  is now given by 
\begin{eqnarray}
  \label{eq:4}
 \Gamma \frac{\partial p(z)}{\partial t} &=& K\frac{\partial^2 p}{\partial z^2}
  -r p + w p^2 - u p^3 -h(z,t) V_1^{\prime}(p),  \nonumber\\  
{\rm with} \ \ 
h(z,t)&=&{\cal U}\left (\frac{z-ct}{w_h}\right ).
\end{eqnarray}
 This is equivalent to adding a term $\int dz h(z,t) V_1(p)$ in the
Hamiltonian, Eq. (\ref{eq:1b}), such that the drive favours the
unzipped region($p=0$) in a region of width $w_h$ around $z=ct$, with
the front position at $t=0$ chosen as origin $z=0$.  In
Eq. (\ref{eq:4}), $ V_1(p)$ should have the right form for
$V(p)+V_1(p)$ to favour the unzipped phase.  Since the helicase works
only near the interface or the front, ${\cal U}(x)$ is a short range
function.  For simplicity, we choose ${\cal U}(x)= \Delta u$ for $\mid
x\mid \le 1$ and zero otherwise.  The ``drive'' is attached to the
front and moves to the zipped side with a speed $c$ which is to be
determined self-consistently so that the front also moves with the
same speed.

The role of the drive term is to  disturb the coexistence
between $p=0$ (unzipped) and $p=p_0$ (zipped phase).   By a
transformation of variables, like $p=p_0-\phi$ and rescaling,
we recast Eq. (\ref{eq:4}) in a more standard and symmetrical form (by
choosing $V_1(p)$)
\begin{eqnarray}
  \label{eq:18}
  \frac{\partial \phi(z)}{\partial t} &=& \frac{\partial^2 \phi}{\partial z^2}
  + f(\phi), \ \ {\rm where} \\ 
f(\phi)&=&\left (- \frac{1}{3} + h(z,t) \right ) \phi +  \phi^2 
   -  \left (\frac{2}{3}+h(z,t)\right )  \phi^3,\nonumber
\end{eqnarray}
and, for brevity, same notation $h$ is used for the drive.  Another
choice could have been $f(\phi)= \phi(1-\phi)(\phi-
\frac{1}{2}-h(z,t))$ which is identical to Eq. (\ref{eq:18}) upto a
scale transformation if and only if $h=$ constant. In any case no
fundamental difference is expected among the various possible
choices.  Eq. (\ref{eq:18}) allows  the zipped phase with 
$\phi=0$ and the unzipped phase with $\phi=1$ for all $h$.  These two
phases coexist at $h=0$, while
$\phi=0$ is metastable for $0<h<1/3$ and unstable for $h>1/3$. See
Fig. \ref{fig:res}. The symmetrical form helps in identifying the
location of the front by $\phi=0.5$ and the drive $h(z,t)$ is
operative only around that point.

Assuming that the front propagates with a velocity $c$,
i.e. $\phi(z,t)=\phi(z-ct)$, We can use the comoving frame with
coordinate $\xi=z-ct$ to rewrite Eq. (\ref{eq:18})as
\begin{equation}
  \label{eq:17} \frac{\partial^2 \phi}{\partial \xi^2}+c\frac{\partial
  \phi(z)}{\partial \xi} +
\frac{\partial}{\partial\phi}\tilde{V}(\phi)=0, \quad {\rm with} \quad
\frac{\partial}{\partial\phi}\tilde{V}(\phi)=-f(\phi)
\end{equation}
which has a mechanical analogy of a particle moving in a potential
$-\tilde{V}(\phi)$ under friction (taking $\xi$ as a time like variable).
This analogy immediately tells us (using first integral or energy
conservation) that to satisfy the boundary conditions at
$\xi=\pm\infty$ when $h=0$, one must have $c=0$.  In other words,
there is no propagating solution as it should be  in the case of
phase coexistence. 
A propagating solution ensues if the drive $h$ is not zero, with
$c=c(\Delta u,w_h)$. 

The speed of propagation $c$ has to be insensitive to width $w_h$ of
the pulse if $w_h >>$ the width of the front or interface.  In that
large $w_h$ limit, $c$ should be the speed of a propagating front
for a uniformly metastable (i.e. $h < 1/3$) or unstable (i.e. $h\ge
1/3$) case. We recollect the relevant results for the uniform
situation with $h(z,t)=$ constant. {\it(i)} There is a pushed to
pulled transition\cite{ebert} in the propagation of the front at
$h=4/3$. {\it(ii)} Beyond $h=4/3$ the velocity is determined by the
linearized equation of motion while the full nonlinearity is important
for $h<4/3$.  {\it(iii)} For the metastable case ($h< 1/3$) any
initial condition $\phi_0(z)=\phi(z,t=0)$ rapidly evolves to a steady
shape with a velocity $c^{\dagger}(h)= 3h/\sqrt{2[(2/3)+h]}$,
approaching the steady speed exponentially fast in time.  {\it(iv)}
For the unstable case, a sharp interface (say a sharp step at $t=0$)
also evolves to this ``pushed'' front so long as $h \le 4/3$.  However a
flatter interface would maintain its flatness and move with a speed
determined by the initial flatness. {\it(v)} In the pulled limit, ($h>
4/3$), the asymptotic speed is the linearized speed
$c^*(h)=2\sqrt{h-(1/3)}$ which is reached algebraically in time
provided the initial condition is sharp (e.g. a step function).

With a pulse, one may use the particle mechanics analogy that a
particle is in one of the peaks of the equal height double peaked
potential $-\tilde{V}(\phi)$ at $\xi=-\infty$ and then at finite time
it gets a push (energy input in particle mechanics) which should be
sufficient to overcome frictional loss and reach the other peak at
$\xi=+\infty$.  This will satisfy the boundary conditions at
$\xi=\pm\infty$.  A moving front is therefore possible.  In other
words the moving front originates from the ``elastic term'' that tries
to spread out the change in $\phi$ in the front region.

\figfour

For a quantitative analysis of the speed, we use a variational
principle\cite{benguria}.  If the equation of motion admits, as we
verify numerically below, a monotonic front $\phi(z,t)=q(\xi)$, then
the inverse mapping can be used to get $\xi$ from $q$ with $0\le q\le
1$.  There is an inequality,
\begin{equation}
  \label{eq:6}
  c^2 \ge 2 \frac{\int_0^1 \ f g dq}{\int_0^1 -(g^2/g^{\prime}) dq}, 
\end{equation}
for any positive function $g(q)$ with $-dg/dq>0$, provided the
integrals exist.  This requires $f^{\prime}(0)<0$ for bistable $f$ of
the form Eq. (\ref{eq:18}) .  Only the metastable case is considered
here.  The unstable case with $f^{\prime}(0) >0$ can also be treated
though in a slightly different way.  Taking $g(q)=[(1-q)/q]^{b}$ and
the uniform case of $f$ i.e. $h(z,t)=h$, one gets
\begin{equation} 
\label{eq:7}
c^2 \ge ( -b^2/3) +  h b(6-b)/2.  
\end{equation} 
The supremum of the lower bound at $b=9h/(2+3h)$ recovers the exact
velocity, $c^{\dagger}$, mentioned earlier (valid for $h<1/3$).  Under
the assumption of monotonicity, the pulse, in $q$-space, is at $q=1/2$
and is of width $\Omega(w_h)$ such that $\Omega(w_h \rightarrow
\infty)=1/2$.  The profile $q(\xi)$ generally approaches the two
limits exponentially as we have seen earlier and therefore
$\tilde{\Omega}\equiv (1/2) - \Omega \sim \exp(-w_h/w^*)$ with some
characteristic length $w^*$.  The pulse term contribution to the
numerator of the bound of Eq. (\ref{eq:6}) is an integral of $q(1-q^2)
g(q)$ over $q \in [\tilde{\Omega}, 1-\tilde{\Omega}]$ and the integral
can be expressed in terms of incomplete beta functions.  We then
obtain
\begin{equation}
  \label{eq:8}
  c^2 \ge \frac{ -b^2}{3} + \frac{1}{2} \Delta u b (6-b) -
  A(\tilde{\Omega},b) \Delta u~ b, 
\end{equation}
where the form of $A(\tilde{\Omega},b) (\ge 0)$ is not displayed. 
Taking the maximum of the right hand side as the best estimate
$c(\Delta u,w_h)$ for the speed, we see that $c(\Delta u,w_h) <c^{\dagger}$ (the bulk
value), as expected.  Using the asymptotic behaviour of the incomplete
beta functions in $A(\tilde{\Omega},b) $, one finds that $c(\Delta u,w)$
saturates exponentially for large $w_h$.  In the other limit of small
$w_h$ (equivalent to small $\Omega$ ), there is a linear dependence on
$w_h$.  Combining these, we may write $c(\Delta u,w)= c_0[1 -a \exp(-w/w_0)]$,
a form that does represent the numerical data very well.

In order to determine the velocity of propagation of the front with a
pulse, we have numerically solved a discretized version of Eq.
(\ref{eq:17}) for various values of the width $w_h$ and magnitude of
the drive $\Delta u$.  A small time step is chosen for proper
convergence but no spatial continuum limit has been done.  We start
with a sharp interface, $\phi=1$ for $0\le z\le L/2$ and $\phi=0$ for
$z> L/2$.  A sequential update is done.  At every time step, we allow
a square pulse of width $w_h$ and strength $\Delta u$ at the current
location of the domain wall or front (located by $\phi=0.5$).  In all
cases we observed a monotonic front.  Fig. \ref{fig:prob} shows the
time evolution of the front for the case of a drive with $\Delta
u=0.2$ and of zero width, $w_h=0$ (pulse at one point of the lattice
only ).

The variation of the speed with the width of the pulse is shown in Fig
\ref{fig:vel}.  Consistent with our results from the variational
analysis, we see that the velocity approaches the bulk limit for large
widths and this approach is exponential.  There is a small but
systematic deviation of the observed velocity from the exponential fit
for larger values of $\Delta u$ (in the ``unstable'' region).
Detailed analysis of the pushed versus pulled cases will be reported
elsewhere.

In terms of helicases, it seems natural to associate dnaB type passive
helicases with the metastable case where the motor action provides the
drive that locally disturbs the Y-fork region.  As in
Ref. \cite{hel2k2}, we identify the domain wall or the front as the
Y-fork region - the junction of the ds-ss DNA.  In the metastable
case, the pushed dynamics has a stability against small fluctuations,
the speed of propagation is independent of the initial conditions, and
the speed approaches the steady state limit exponentially in time.
All of these are important properties expected of a helicase of type
dnaB which, after loading on DNA, carries out the unzipping in tandem
with the other processes during replication.

\figfive

So far as PcrA (Fig. \ref{fig:hel}) is concerned, we associate the
overall dynamics to the unstable case.  No quantitative experimental
results are available regarding the magnitude and width of the force
PcrA exerts on the bases beyond the Y-fork.  It is reasonable to
assume that the force is meant to unzip DNA locally, and the effect of
this pulling is to make the ds-region unstable.  We then infer that
PcrA operates in the unstable ($h>1/3$) regime.  In several mutants of
PcrA (replacing a few residues by alanines) the helicase activity (the
hand in Fig \ref{fig:hel}) could be decoupled (reduced) from the motor
action and ATP intake (both remained more or less the same).  In our
terminology, these mutations involve a reduction in $h$ (i.e.  the
overall pulling strength, $\Delta u$, or the width of the pulse,
$w_h$, or both), producing a reduction in the speed as shown in
Fig. \ref{fig:vel}.  We like to add that reversed motion of recG can
also be understood in the same scheme with a few extra
ingredients. This will be discussed elsewhere.  Active helicases like
PcrA that are involved in repair processes become operational when the
replication process stalls because of, e.g., defects.  Such a stalling
would lead to a relaxation of the stalled front.  The new relaxed
$\phi(z)$ would then act as the initial condition for the new helicase
recruited for repair.  The sensitivity to initial condition of a front
invading an unstable phase is an important distinction between the
metastable (pushed) and the unstable cases.  Whether helicases in
charge of repair are really sensitive to and recognize these initial
conditions need to be probed experimentally.

To summarize, we have proposed a simple coarse grained model for
describing helicase activity on DNA.  The DNA is described by a local
probability of zipping of the base pairs.  The motor action of the
helicase induces an instability around the front (Y-fork) in an
otherwise coexisting zipped and unzipped phases.  Such coexisting
phases with an interface can in principle be created by pulling or by
a stalled helicase on a dsDNA and therefore can be studied
experimentally.  We have shown that the local drive created by the
helicase leads to a traveling wave solution with a selected velocity
that depends on the nature of the drive.  Quantitative studies of
forces and sensitivity to the initial conditions would provide crucial
clues on the nature of dynamics studied in this paper.  We hope single
molecular experiments in future would be able to probe these in
detail.

\end{document}